\documentclass[conference, doublecolumn, hidelinks, xcolor=dvipsnames]{IEEEtran}

\pdfoutput=1
\IEEEoverridecommandlockouts
\usepackage{microtype}
\usepackage{acronym}
\usepackage{amsmath}
\usepackage{amssymb}
\usepackage[english]{babel}
\usepackage{cite}
\usepackage[shortlabels,inline]{enumitem}
\usepackage{graphicx}
\usepackage{hyperref}
\usepackage{cleveref}
\usepackage[utf8]{inputenc}
\usepackage{lipsum}
\usepackage{siunitx}
\usepackage[dvipsnames]{xcolor}
\usepackage[T1]{fontenc}
\usepackage{threeparttable}

\usepackage{bbm}

\newtoggle{Arxiv}

\toggletrue{Arxiv}

\newcommand{\version}[2]{%
\iftoggle{Arxiv}{%
#1%
}%
{#2}%
}%%
\allowdisplaybreaks

\newcommand{\Equation}[1]{\eqref{#1}}

\newcommand{\Section}[1]{Section~\ref{#1}}

\renewcommand{\vec}[1]{\mathbf{#1}}

\renewcommand{\Pr}[1]{\mathrm{Pr}\left\lbrace #1 \right\rbrace}
\newcommand{\CPr}[2]{\mathrm{Pr}\left\lbrace #1 \,|\, #2 \right\rbrace}
\newcommand{\InF}[2]{\mathbbm{1}_{#1}\left(#2\right)}
\newcommand{\InFt}[2]{\tilde{\mathbbm{1}}_{#1}\left(#2\right)}

\newcommand{\sysX}{\mathcal{X}}
\newcommand{\envY}{\mathcal{Y}}
\newcommand{\statesX}{\mathcal{A}_{\mathcal{X}}}
\newcommand{\statesY}{\mathcal{A}_{\mathcal{Y}}}
\newcommand{\statesYnorm}{\mathcal{A}_{\mathcal{\tilde{Y}}}}
\acrodef{ABM}{agent-based modeling}
\acrodef{AoI}{age of information}
\acrodef{BER}{bit error rate}
\acrodef{CAS}{complex adaptive system}
\acrodef{CB}{chemotactic bacterium}
\acrodef{MC}{molecular communication}
\acrodef{pmf}{probability mass function}

\makeatletter
\long\def\@makecaption#1#2{\ifx\@captype\@IEEEtablestring%
    \footnotesize\begin{center}{\normalfont\footnotesize #1}\\
        {\normalfont\footnotesize\scshape #2}\end{center}%
    \@IEEEtablecaptionsepspace
    \else
    \@IEEEfigurecaptionsepspace
    \setbox\@tempboxa\hbox{\normalfont\footnotesize {#1.}~~ #2}%
    \ifdim \wd\@tempboxa >\hsize%
    \setbox\@tempboxa\hbox{\normalfont\footnotesize {#1.}~~ }%
    \parbox[t]{\hsize}{\normalfont\footnotesize \noindent\unhbox\@tempboxa#2}%
    \else
    \hbox to\hsize{\normalfont\footnotesize\hfil\box\@tempboxa\hfil}\fi\fi}
\makeatother

\newcommand{\scaleSection}{\vspace{-0.2cm}}
\newcommand{\scaleSubsection}{\vspace{-0.12cm}}
\newcommand{\scaleSubsubsection}{\vspace{-0.0cm}}
\newcommand{\scaleSectionBelow}{\vspace{-0.1cm}}
\newcommand{\scaleSubsectionBelow}{\vspace{-0.0cm}}
\newcommand{\scaleSubsubsectionBelow}{\vspace{-0.0cm}}
\newcommand{\scaleAlign}{\vspace{-0.0cm}}

\begin{document}
\bstctlcite{IEEEexample:BSTcontrol}
\title{Semantic Information in MC: Chemotaxis Beyond Shannon\vspace{-0.3cm}}
\author{
\IEEEauthorblockN{Lukas Brand\IEEEauthorrefmark{1}, Yan Wang\IEEEauthorrefmark{1}, Maurizio Magarini\IEEEauthorrefmark{2}, Robert Schober\IEEEauthorrefmark{1}, and Sebastian Lotter\IEEEauthorrefmark{1}}\\[-0.2cm]
\IEEEauthorblockA{\small\IEEEauthorrefmark{1}Friedrich-Alexander-Universit\"at Erlangen-N\"urnberg (FAU), Erlangen, Germany}\\
\vspace{-0.3cm}
\IEEEauthorblockA{\small\IEEEauthorrefmark{2}Polytechnic University of Milan (POLIMI), Milan, Italy}\\
\vspace{-1.1cm}
}
\maketitle
\thispagestyle{plain}
\pagestyle{plain}
\begin{abstract}
The recently emerged \ac{MC} paradigm intends to leverage communication engineering tools for the design of synthetic chemical communication systems.
These systems are envisioned to operate at nanoscale and in biological environments, such as the human body, and catalyze the emergence of revolutionary applications in the context of early disease monitoring and drug targeting.
Despite the abundance of theoretical (and recently also experimental) \ac{MC} system designs proposed over the past years, some fundamental questions remain unresolved, hindering the breakthrough of \ac{MC} in real-world applications.
One of these questions is: What can be a useful measure of {\em information} in the context of \ac{MC} applications?
While most existing works on \ac{MC} build upon the concept of {\em syntactic information} as introduced by Shannon, in this paper, we explore the framework of {\em semantic information} as introduced by Kolchinsky and Wolpert for the information-theoretic analysis of a natural \ac{MC} system, namely bacterial chemotaxis.
Exploiting computational \ac{ABM}, we are able to quantify, for the first time, the amount of information that the considered \ac{CB} utilizes to adapt to and survive in a dynamic environment.
In other words, we show how the flow of information between the environment and the \ac{CB} is related to the \textit{effectiveness} of communication.
Effectiveness here refers to the adaptation of the \ac{CB} to the dynamic environment in order to ensure survival.
Our analysis reveals that it highly depends on the environmental conditions how much information the \ac{CB} can effectively utilize for improving their survival chances.
Encouraged by our results, we envision that the proposed semantic information framework can open new avenues for the development of theoretical and experimental \ac{MC} system designs for future nanoscale applications.

\end{abstract}
\setlength{\belowdisplayskip}{2pt}
\setlength{\belowdisplayshortskip}{2pt}
\acresetall
\scaleSection
\section{Introduction }\label{sec:intro}
\scaleSectionBelow
Transmission of information occurs at several levels of abstraction and, according to Weaver, can be categorized into the communication of (i) accurate syntactic information (technical level), (ii) the meaning/significance of a message (semantic level), and (iii) the actions caused by a message (effectiveness level) \cite{weaver1953recent}.
While unappreciated for a long time, the study of the semantic and effectiveness levels (the latter one also referred to as the goal-oriented level) has recently received significant attention in the context of future sixth-generation (6G) communication networks; for example, for communication in personalized body area networks, unmanned aerial vehicles-assisted networks, and networks of autonomous collaborative robots \cite{shi2021semantic,yang2022semantic, strinati20216g}.

In this paper, we do not consider conventional communication systems based on electromagnetic waves, but the recently emerged \ac{MC} systems; in these systems, information is transmitted by chemical messengers.
Due to the physical nature of molecular transport, synthetic \ac{MC} systems are, according to experimental and theoretical studies, characterized by relatively poor performance on the syntactic level (low achievable data rates, high bit error rates), whereas extremely good performance is often observed on the effectiveness level in natural \ac{MC} systems (e.g., complex adaptations of single-cell species to changing environmental conditions); this observation indicates that semantic information may be a highly relevant measure to consider in the analysis of \ac{MC} systems.
In fact, a shift in perspective towards the analysis of semantic information and the respective communication goal (effectiveness) may be beneficial for guiding the design of future synthetic \ac{MC} systems and their applications.
Here, we explore this idea by developing a semantic information framework for bacterial chemotaxis, a widely considered natural \ac{MC} system which has already inspired several disruptive synthetic \ac{MC} applications such as collaborative nanodevice-assisted early disease detection and targeted drug delivery \cite{akyildiz2015internet}.

There have been only a few previous studies aiming at creating an information-theoretic framework for \ac{MC} based on semantic communication \cite{ruzzante2023synthetic, barker2023metric, sowinski2023semantic}.
All these works are inspired by the framework of semantic information as introduced by Kolchinsky and Wolpert \cite{kolchinsky2018semantic}.
In \cite{ruzzante2023synthetic}, semantic information was proposed for the design of synthetic cells.
In \cite{barker2023metric}, the concept of {\em subjective information} was introduced in order to measure how much useful information was obtained under different information acquisition strategies relative to a default strategy.
Here, the source of information was an environmental signal, and the useful information was defined as the mutual information between the subject's \textit{actions} and the environmental signal.
However, both studies \cite{ruzzante2023synthetic, barker2023metric} do not consider the dynamic information exchange between the cell/the subject and its environment.
Furthermore, the model systems studied in \cite{ruzzante2023synthetic} and \cite{barker2023metric} were developed specifically for the respective information-theoretic analysis and, consequently have very limited complexity; hence, it is not clear to what extent they generalize to practical real-world systems.
Finally, in \cite{sowinski2023semantic}, the authors present an analysis of the semantic information of a resource gathering agent, which is similar to a bacterium performing chemotaxis.
However, the bacterial propagation model considered in \cite{sowinski2023semantic} is oversimplified and assumes that the location of the nearest resource is always perfectly known to the agent.

In contrast to existing works, in this paper, we apply the generic framework proposed in \cite{kolchinsky2018semantic} to a established computational \textit{run-and-tumble} chemotaxis model; in this way, we are able to demonstrate the applicability of the framework to a practically relevant \ac{MC} system for the first time.
Furthermore, we quantify the adaptability of the considered bacteria to their dynamic environment.
Thus, the framework presented in this paper measures for the first time the dynamic transmission of information on the semantic level in the context of \ac{MC}.
Finally, the presented semantic information analysis is linked to the survival probability of the bacteria, i.e., the effectiveness level of the considered \ac{MC} system.

The remainder of this paper is organized as follows.
Sections~\ref{sec:system_model} and \ref{sec:it_framework} introduce the considered bacterial chemotaxis model and the proposed semantic information framework, respectively.
Numerical results are presented in \Section{sec:evaluation}, before \Section{sec:conclusion} concludes the paper and outlines topics for future work.

\scaleSection
\section{System Model}\label{sec:system_model}
\scaleSectionBelow
\subsection{Bacterial Chemotaxis: Gradient-based Nutrient Detection}\label{sec:chemotaxis}
\scaleSubsectionBelow

Bacterial chemotaxis refers to the ability of some bacteria, called {\em chemotactic bacteria (\acsp{CB}\acused{CB})}, to detect and follow a gradient of attractants, e.g., nutrients, in their environment\footnote{Some \acp{CB} follow the {\em negative} gradient instead of the positive to avoid contact with certain repellents, e.g., poisonous substances. In this paper, however, we focus on nutrient-seeking \acp{CB}.}.
Specifically, as observed in real-world experiments, \acp{CB} employ cell-surface receptors that interact chemically (bind/unbind) with attractants in the environment to sense the attractants' local concentration \cite{berg2004coli}.
Then, based on whether or not a bacterium is moving towards increasing nutrient concentrations, i.e., the sensed concentration exceeds the previously sensed attractant concentration, it keeps moving in the same direction ({\em run mode}) or starts moving randomly until it detects an increasing gradient ({\em tumble mode}).
The \ac{CB}'s run and tumble modes are caused by the counterclockwise and clockwise rotation of its flagellum, respectively, which leads to straight (run mode) or random (tumble mode) movement.
The switching between run mode and tumble mode is achieved by intracellular chemical signaling pathways that are activated in response to the binding/unbinding of the \ac{CB}'s cell-surface receptors.

\begin{figure}[!tbp]
\centering
  \includegraphics[width=0.95\columnwidth]{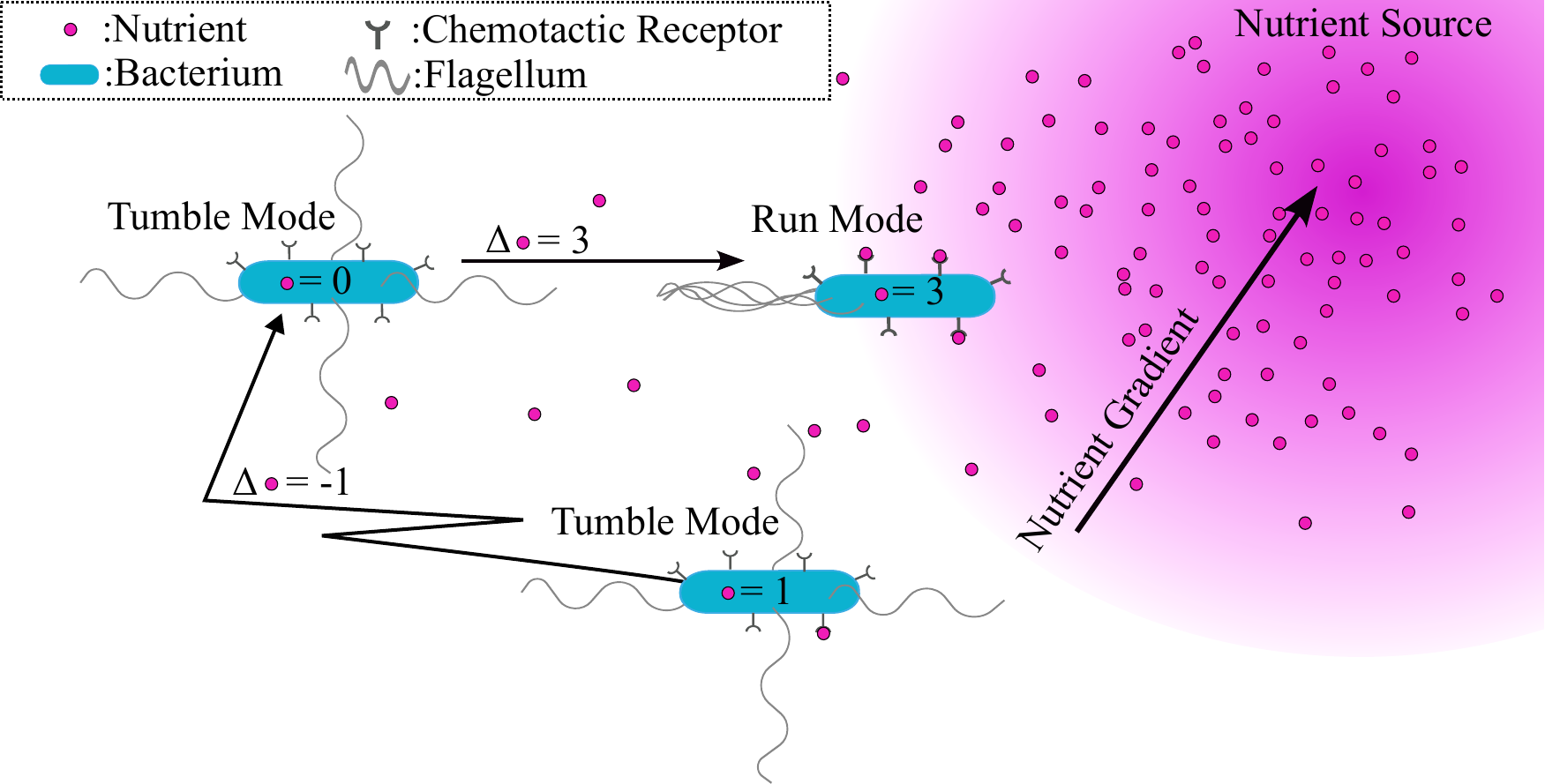}\vspace*{-0.15cm}
  \caption{Schematic illustration of a \ac{CB} (cyan ellipsoid) at three different time instances performing gradient sensing to locate a nutrient source (purple area) that is constantly releasing nutrients (purple dots). The \ac{CB} senses its environment and switches from tumble mode to run mode to follow a positive nutrient gradient, while switching back to tumble mode when the sensed number of nutrients decreases. }
  \label{fig:chemotaxis}
  \vspace*{-0.3cm}
\end{figure}

Fig.~\ref{fig:chemotaxis} illustrates the concept of bacterial chemotaxis schematically.
In particular, the figure shows the same \ac{CB} (cyan ellipsoid) at three different time instants seeking nutrients that are produced by a nutrient source in the top right corner.
At the first time instant (center bottom), only one of the \ac{CB}'s receptors is bound by a nutrient (purple dot) and it moves in tumble mode.
After a random movement to the top left, the \ac{CB} finds itself in a local environment with even fewer nutrients than before.
Consequently, none of its receptors is bound at the second time instant, i.e., the detected nutrient gradient in the direction of movement is negative, and the \ac{CB} remains in tumble mode.
Finally, after (coincidentally) moving in the direction of the food source in the third time instant, the number of bound receptors exceeds the number of previously bound ones and the \ac{CB} switches to run mode, aligning its future trajectory with the location of the nutrient source.

Despite its conceptual simplicity, bacterial chemotaxis presents an example of a {\em \ac{CAS}} \cite{nagarajan2022agent}; the \ac{CB} adapts to a changing environment based on the information it gathers from its surrounding, seeking to maximize its chances to survive by collecting nutrients \cite{berg2004coli}.
On both a practical and a more abstract level, this observation relates the natural bacterial chemotactic system to the synthetic \ac{MC} systems envisioned to operate in biological environments fulfilling complex tasks like drug delivery or early disease diagnosis.
On a practical level, the ability to detect and follow a gradient of molecules in the environment that \acp{CB} possess, is desirable also for many synthetic applications.
For example, autonomous drug-carrying nanodevices could be engineered to follow a gradient of biomarkers in order to deposit their therapeutic charge close to the diseased tissue.
On a more abstract level, for many envisioned \ac{MC} applications it is still an open question how information exchange (for example between nanodevices or between nanodevices and their environment) relates {\em quantitatively} to achieving a particular goal, e.g., cooperative early disease diagnosis or drug delivery.
Since information acquisition (nutrient sensing) and goal (survival) are well-characterized in the context of bacterial chemotaxis, this provides an opportunity for developing and testing such a quantitative information-theoretic framework.

As a playground for exploring these possibilities, we will introduce an existing computational bacterial chemotaxis model in the remainder of this section that we will utilize later to develop and test the semantic information theory framework proposed in this paper.

\scaleSubsection
\subsection{Agent-based Model for Bacterial Chemotaxis}\label{sec:abm}
\scaleSubsectionBelow

\Acp{CAS}, such as the bacterial chemotaxis studied in this paper, are notoriously difficult to formalize and analyze mathematically; mainly because the complex nonlinear dynamics of these systems usually emerge from (local) interactions among discrete entities with large state-spaces which are difficult or impossible to capture and analyze using analytical frameworks.
However, since \acp{CAS} arise in many and diverse scientific fields, ranging from economy over society to systems biology, the need for modeling their dynamics has spawned a number of innovative computational modeling tools, among which \ac{ABM} is a particularly commonly used one \cite{abar2017agent}.
In \acp{ABM}, computational agents interact with each other and/or the environment according to a set of pre-defined rules.
The rules specify which action an agent is going to perform based on some information that it collects from other agents and/or the environment.
\version{In econometric models, the agents could, for example, model stock traders that interact with the stock market (environment) and other traders according to certain buying and selling strategies.
The particular benefit of \acp{ABM} for the analysis of \acp{CAS} is their exploratory power; in the example above, they can, for example, be utilized to test the impact of different trading strategies on the stock market dynamics.}{}

\begin{figure}[!tbp]
\centering
  \includegraphics[width=0.95\columnwidth]{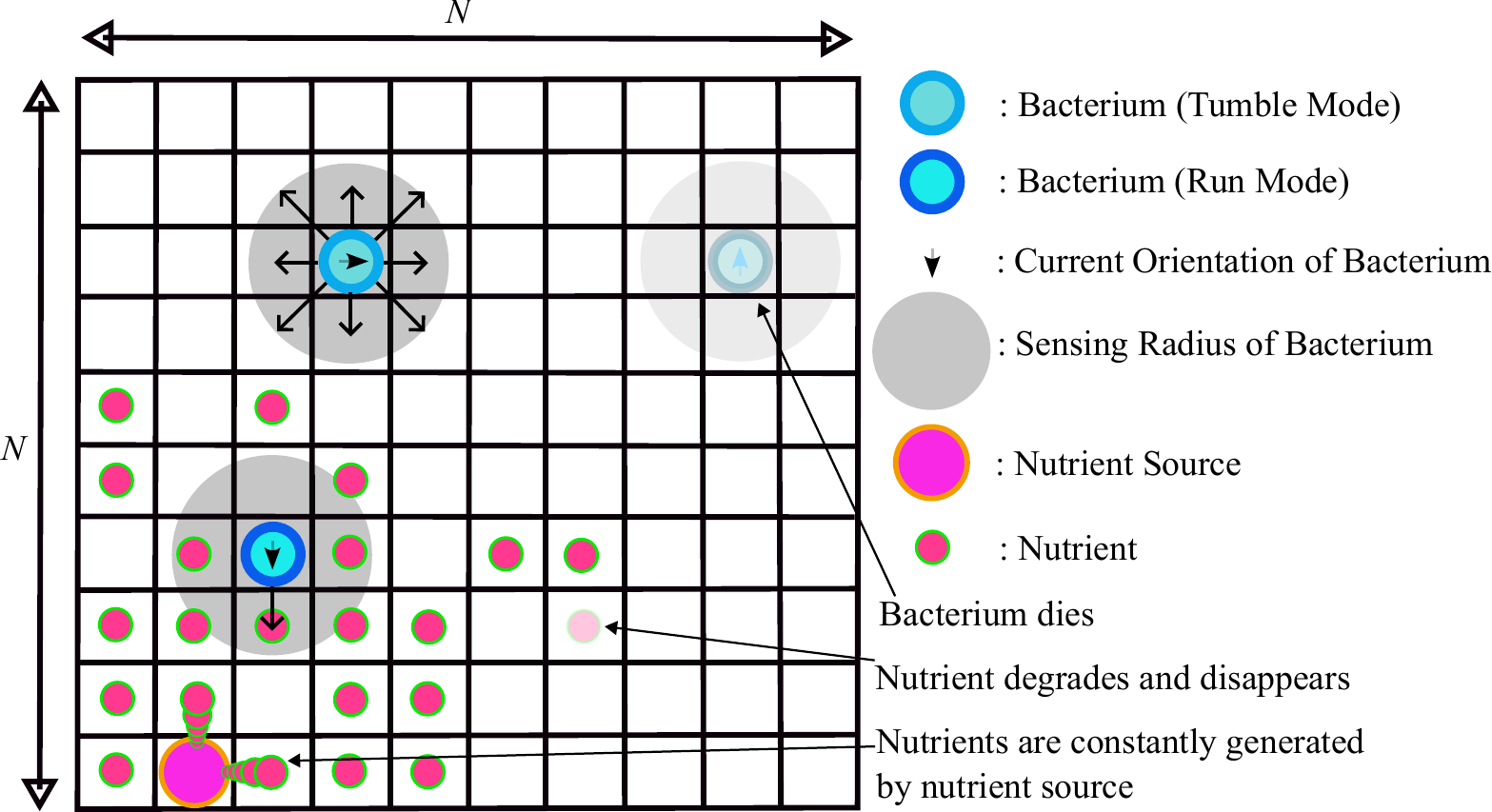}\vspace*{-0.15cm}
  \caption{Considered computational bacterial chemotaxis model. Bacterium (cyan) seeks to follow nutrient source (pink) by switching between run and tumble modes according to the nutrient (red) concentration it observes in its local environment.}
  \label{fig:abm}
  \vspace*{-0.2cm}
\end{figure}

In the following, we introduce an \ac{ABM} for bacterial \textit{run-and-tumble}-based chemotaxis which is adopted from the literature \cite{nagarajan2022agent, berg2004coli} and simplified for our purposes.
Fig.~\ref{fig:abm} illustrates the main components of the considered model.
As computational domain, it features a two-dimensional lattice of size $N$-by-$N$ denoted by $\mathcal{L} = \lbrace 1,\ldots, N \rbrace \times \lbrace 1,\ldots, N \rbrace$ on which a \ac{CB} (cyan) tries to locate nutrients (red) which are supplied by a nutrient source (pink).
The state of the computational model evolves in discrete time steps $k \in \mathbb{N}_0$, where $\mathbb{N}_0$ denotes the set of non-negative integers, as a result of the actions and interactions of the components described in the following.

\scaleSubsubsection
\subsubsection{Nutrient Source}\label{sec:nutrient_source}
\scaleSubsubsectionBelow

The \ac{ABM} features one {\em nutrient source} whose purpose is to supply the environment with nutrients.
The location of the nutrient source in time step $k$ is denoted by $\vec{s}[k] \in \mathcal{L}$.
In each time step, the nutrient source produces nutrients at rate $k_{\mathrm{NS}}$ [nutrients/time step] which are placed randomly on one of the neighboring grid cells (cells at distance $1$ from $\vec{s}[k]$), where the distance between two grid cells $\vec{c}_1 = [i_1,j_1] \in \mathcal{L}$ and $\vec{c}_2 = [i_2,j_2] \in \mathcal{L}$ is defined as $d(\vec{c}_1, \vec{c}_2) = \max\lbrace |i_1-i_2|, |j_1-j_2| \rbrace$. See also Fig.~\ref{fig:abm} for an illustration of this process.
In addition to producing nutrients, the nutrient source is mobile and changes its location randomly every $m$ time steps to a new position $\vec{s}[k+m]$ at a distance $d_{\mathrm{max}}$, i.e., $d(\vec{s}[k], \vec{s}[k+m]) = d_{\mathrm{max}}$.
The mobility of the nutrient source resembles the variability of the food supply in natural systems and challenges the food-seeking bacterium to follow its trajectory; the more challenging the access to a nutrient source, here represented by a higher mobility of the nutrient source, the more sophisticated are the adaptation strategies that the bacterium needs \cite{hibbing2010bacterial}.

\scaleSubsubsection
\subsubsection{Nutrients}\label{sec:nutrient}
\scaleSubsubsectionBelow

The nutrients produced by the nutrient source remain at their spawning positions and vanish randomly after a minimum number of time steps $K_{\mathrm{deg}} = 10$.
The vanishing of the nutrients from the grid resembles the degradation of nutrients in biological systems.
Let $N_Y[k]$ denote the number of nutrients at time step $k$.
Then, the positions of all nutrients at $k$ are collected in vector $\vec{Y}[k] = [\vec{Y}_1^{(k)},\ldots,\vec{Y}_{N_Y[k]}^{(k)}] \in \mathcal{L}^{N_Y[k]}$, where $\vec{Y}_i^{(k)} \in \mathcal{L}$ for $i \in \lbrace 1,\ldots, N_Y[k] \rbrace$.

\scaleSubsubsection
\subsubsection{Bacteria}\label{sec:bacteria}
\scaleSubsubsectionBelow
Similar to the biological \acp{CB} discussed at the beginning of this section, the computational \ac{CB} (agent) scans its local environment for food and adapts its movement strategy according to the sensed nutrient concentration\footnote{Due to space constraints, we consider here a system with a single \ac{CB}. The evaluation of a system with more than one \ac{CB}, which would allow the investigation of possible interactions between the \acp{CB}, such as competition for resources, is left for future work.}.
Let us denote the position of the \ac{CB} at time step $k$ by $\vec{x}[k] \in \mathcal{L}$.
Then, the \ac{CB} counts the number of nutrients $N^{\mathrm{o}}[k]$ in its immediate vicinity, i.e., on its own grid cell and on the neighboring grid cells, as $N^{\mathrm{o}}[k] = |\lbrace \vec{Y}_l^{(k)} \,|\, d(\vec{Y}_l^{(k)}, \vec{x}[k]) \leq 1, 1 \leq l \leq N_Y[k]  \rbrace|$, where $|\mathcal{A}|$ denotes the cardinality of the set $\mathcal{A}$. Thus, the \ac{CB} senses a nutrient concentration $N^{\mathrm{o}}[k]$ ranging between $0$ and $9$ at each time step $k$.
Finally, as illustrated in Fig.~\ref{fig:abm}, the \ac{CB} traverses $\mathcal{L}$ in one of two modes, the tumble mode or the run mode, resembling the two operating modes of biological \acp{CB} as discussed above.
It switches into the tumble mode in time step $k$ if $N^{\mathrm{o}}[k] < N^{\mathrm{o}}[k-1]$ or $N^{\mathrm{o}}[k] = 0$ and into the run mode otherwise.
In tumble mode, $\vec{x}[k+1] \in \mathcal{L}$ is selected randomly such that $d(\vec{x}[k+1], \vec{x}[k]) = 1$, i.e., the \ac{CB} randomly moves to one of the neighboring grid cells.
In run mode, $\vec{x}[k+1] = [\vec{x}[k] + (\vec{x}[k] - \vec{x}[k-1])]_{\mathcal{L}}$, where the notation $[\vec{x}]_{\mathcal{L}}$ indicates that vector $\vec{x}$ is mapped towards the nearest location inside $\mathcal{L}$; hence, in run mode, the \ac{CB} continues its previous movement direction straight unless it hits a boundary from which it is reflected.
If, due to a reflection at the boundary of $\mathcal{L}$, $\vec{x}[k] = \vec{x}[k-1]$, the \ac{CB} resets its current direction to a new random direction, i.e., in the next time step, $\vec{x}[k+1] \in \mathcal{L}$ is selected randomly such that $d(\vec{x}[k+1], \vec{x}[k]) \leq 1$.
The \ac{CB} also possesses a {\em weight}, which is initially $1$, and decreases by an amount of $0.05$ in each time step. Moreover, the weight increases by a constant amount of $0.2$ (up to a maximum value of $1$) each time $\vec{x}[k]$ coincides with one of the nutrient positions $\vec{Y}[k]$.
When the weight of the \ac{CB} reaches $0$, the \ac{CB} dies and can no longer move.

We note that the computational \ac{CB} model considered here, although rather simple, can be easily extended in future work by considering, e.g., (i) noisy intra-cellular signal propagation, (ii) memory in the \ac{CB}'s decision making \cite{gosztolai2020cellular}, (iii) the energy costs to acquire and store information \cite{tjalma2023trade}, (iv) more sophisticated gradient detection methods, e.g., the spatial gradient sampling method from \cite{rode2023chemotactic}, where the receptor position and state (bound or unbound) are utilized for aligning the \ac{CB}'s movement direction with the location of the nutrient source, and/or (v) \textit{infotaxis} detection methods, where the search trajectories feature ‘zigzagging’ paths \cite{vergassola2007infotaxis}.
In fact, the semantic information theory framework presented in the following section even allows for the comparative analysis of these and further \ac{CB} models, since it is based directly only on the observed behavior of the \ac{CB} and only indirectly on its specific food-seeking strategy.

\scaleSection
\section{Semantic Information Theory Framework}\label{sec:it_framework}
\scaleSectionBelow
In this section, we briefly outline the general semantic information framework for physical non-equilibrium systems proposed in \cite{kolchinsky2018semantic} and then specialize this framework to the considered computational bacterial chemotaxis model.

\scaleSubsection
\subsection{Semantic Information in Bio-Physical Systems}\label{sec:syn_vs_sem}
\scaleSubsectionBelow

To motivate the following discussion and definitions, let us consider first on an abstract level what is the meaning (or significance) of information for \acp{CB}.
Ultimately, the {\em goal} of information acquisition (cf.~discussion of the effectiveness level of communication in Section~\ref{sec:intro}) for a \ac{CB} is to choose its actions in line with its own state and the state of the environment in order to maximize its survival chances.
Hereby, the information/knowledge about the state of the environment is acquired by the \ac{CB} through the sensing process.
However, perfect knowledge about the state of the environment is not needed {\em for the life-sustaining action}, i.e., not all of the information about the environment is equally important. In fact, from all the information that the \ac{CB} gathers about its environment, only that part carries significance which helps the \ac{CB} to align its actions with its own state and the state of its environment.
The semantic information framework outlined in the following builds upon this fact in the sense that it measures only that information that actually carries significance in the above sense.

Rooted in the theory of dynamical physical systems, the authors in \cite{kolchinsky2018semantic} propose the {\em transfer entropy} between a dynamical system $\sysX$ and its dynamical environment $\envY$ as a sensible measure of the degree to which $\sysX$ is able to adapt to $\envY$ as both change over time.
In this context, $\sysX$ and $\envY$ are characterized at any time $k$ by their {\em states} $X_k \in \statesX$ and $Y_k \in \statesY$, respectively; under the assumption that {\em state spaces} $\statesX$ and $\statesY$ are discrete, the joint dynamics of $\sysX$ and $\envY$ are then defined by the {\em transition probabilities} $\CPr{X_{k+1}, Y_{k+1}}{X_{k}, Y_{k}}$ and the {\em initial distribution} $\Pr{X_{0}, Y_{0}}$.
Then, the transfer entropy between $\sysX$ and $\envY$ accumulated over $k$ time instances is defined as \cite{kolchinsky2018semantic}
\scaleAlign
\begin{equation}\label{eq:it:transfer_entropy}
    \mathcal{T}[k] = \sum_{k'=0}^{k-1} I(X_{k'+1};Y_{k'} | X_{k'}),
\end{equation}
where $I(X_{k+1};Y_{k} | X_{k})$ denotes the {\em conditional mutual information} between the system's future state $X_{k+1}$ and the present state of the environment $Y_{k}$ given the present state of the system $X_{k}$.
In \eqref{eq:it:transfer_entropy}, $I(X_{k+1};Y_{k} | X_{k})$ is computed according to its definition in \cite[Eq. (8.10)]{Mackay2003information} using the conditional \ac{pmf} $p_k(x_2\,|\,x_1,y_1) = \CPr{ X_{k+1}=x_2}{X_{k}=x_1, Y_{k}=y_1}$, where $x_1,x_2 \in \statesX$ and $y_1 \in \statesY$, and the marginal probabilities of $Y_k$, $\Pr{Y_k}$.

We note that $p_k(x_2\,|\,x_1,y_1)$ results after separating the joint system-environment dynamics as follows
\scaleAlign
\begin{align}\label{eq:it:sys_dyn}
 &\CPr{X_{k+1}, Y_{k+1}}{X_{k}, Y_{k}}\nonumber\\
 &\quad= \CPr{Y_{k+1}}{X_{k}, X_{k+1}, Y_{k}} \, \CPr{X_{k+1}}{X_{k}, Y_{k}}.
\end{align}
Introducing the short-hand notation $q_k(y_2\,|\,x_1,x_2,y_1) = \CPr{ Y_{k+1}=y_2}{X_{k}=x_1, X_{k+1}=x_2, Y_{k}=y_1}$, $y_2 \in \statesY$, \eqref{eq:it:sys_dyn} suggests that the system dynamics can be separated as $q_k \, p_k$ into the state evolution of the environment, $q_k$, and the action/state evolution of the system, $p_k$.
Here, the exact forms of $p_k$ and $q_k$ depend on (a) the specific choices of $\sysX$ and $\envY$ and (b) how $\sysX$ and $\envY$ interact, i.e., how they mutually influence the dynamics of each other.
In particular, $p_k$ is defined by what information the dynamical system $\sysX$ gathers about $Y_k$ and how it selects its actions accordingly.

Now, the key idea for quantifying semantic information as proposed in \cite{kolchinsky2018semantic} is to consider different {\em interventions} that change the way that $\sysX$ perceives (and/or reacts to) $Y_k$ leading to different system dynamics $p^{\iota}_k(x_2\,|\,x_1,y_1)$, where $\iota \in \mathcal{I}$ denotes such an intervention from the set of all possible interventions $\mathcal{I}$.
One simple example for an intervention is that $\sysX$ may not be able to distinguish among a set of different environment states under the intervention, while it is able to distinguish these states under normal conditions.
Then, for a given intervention $\iota$, the corresponding transfer entropy is defined as
\scaleAlign
\begin{equation}
    \mathcal{T}_{\iota}[k] = \sum_{k'=0}^{k-1} I_{\iota}(X_{k'+1};Y_{k'} | X_{k'}),
\end{equation}
where the conditional mutual information $I_{\iota}(X_{k+1};Y_{k} | X_{k})$ is computed based on $p^{\iota}_k(x_2\,|\,x_1,y_1)$ and the marginal distribution of $Y_k$ under the {\em intervened system dynamics} $q_k p^{\iota}_k$.

In addition to quantifying the degree of interaction between the dynamical system $\sysX$ and the dynamical environment $\envY$ via the transfer entropy, the concept of semantic information considered in \cite{kolchinsky2018semantic} necessitates a scalar-valued {\em viability function} $V[k]$ that provides a measure of the level of existence (fitness) of $\sysX$ at time step $k$. A specific viability function for the bacterial chemotaxis scenario is defined in the next section.
Considering different interventions $\iota \in \mathcal{I}$, we expect that $V[k]$ can be different under some intervened system dynamics as compared to the normal system dynamics, since some interventions may reduce the ability of $\sysX$ to achieve a particular viability.
Hence, we denote by $V_{\iota}[k]$ the viability of $\sysX$ at time step $k$ under intervention $\iota$.

From the above definitions finally arises the measure of {\em observed semantic information} at time $k$ as the lowest amount of transfer entropy (obtained under the set of interventions $\mathcal{I}$) at which $V_{\iota}[k]$ is not significantly reduced as compared to the original dynamics of the system.
Formally, the observed semantic information at time step $k$ is defined as
\scaleAlign
\begin{equation}\label{eq:it:sem_inf}
    S_{\epsilon}[k] = \min\limits_{\iota \in \mathcal{I}} \left\lbrace \mathcal{T}_{\iota}[k] \,|\, V_{\iota}[k] \geq V[k] (1-\epsilon) \right\rbrace,
\end{equation}
where we have slightly generalized the definition of observed semantic information from \cite{kolchinsky2018semantic}. Our definition allows for a small reduction in viability relative to $V[k]$ by $\epsilon \geq 0$, which accounts for variations of the viability that are negligible in practice; the exact value of $\epsilon$ is determined by the goal of the specific considered \ac{MC} system.

We will show next in the context of the computational \acp{CB} introduced in Section~\ref{sec:system_model} how the above abstract definitions specialize for specific choices of $\sysX$ and $\envY$.

\scaleSubsection
\subsection{Semantic Information for Chemotaxis}\label{sec:sem_in_chemotaxis}
\scaleSubsectionBelow

For the computational chemotaxis system introduced in Section~\ref{sec:abm}, we first specialize $\sysX$ as the computational \ac{CB} and $\envY$ as the nutrients, respectively.
Accordingly, we define the {\em state of the \ac{CB}} at time $k$ as its position $\vec{x}[k]$ and the {\em state of the environment} as the positions of all nutrients $\vec{Y}[k]$, i.e., $\statesX = \mathcal{L}$ and $\statesY = \bigcup_{l = 0,\ldots,N^2} \mathcal{L}^l$, where the definition of $\statesY$ results from the fact that any number of nutrients between $0$ and $N^2$ can exist at the same time and the empty set $\mathcal{L}^0 = \lbrace\rbrace$ corresponds to the state in which no nutrients exist.

In the computational chemotaxis model, the system dynamics $q_k \, p_k$ arise from the algorithmic specifications of the \ac{CB}, the nutrient source, and the nutrients.
Hence, we estimate $q_k \, p_k$ by running ensemble simulations.
Let $E$ denote the total number of simulation runs and let $\vec{x}^e[k]$ and $\vec{Y}^e[k]$ denote the \ac{CB}'s position and the nutrients' positions, respectively, at time $k$ in simulation run $e \in \lbrace 1,\ldots,E\rbrace$.
Then, $q_k \, p_k$ can in principle be obtained from
\scaleAlign
\begin{align}\label{eq:it:ens_avg}
    &\Pr{X_{k+1}=x_2, Y_{k+1}=y_2,X_{k}=x_1, Y_{k}=y_1}\nonumber\\
    &= \lim\limits_{E \to \infty} \frac{1}{E} \sum\limits_{e=1}^{E} \InF{\{x_1\}}{\vec{x}^e[k]} \InF{\{y_1\}}{\vec{Y}^e[k]}\nonumber\\
    &\quad{}\times \InF{\{x_2\}}{\vec{x}^e[k+1]} \InF{\{y_2\}}{\vec{Y}^e[k+1]},
\end{align}
where $\InF{\mathcal{A}}{x}$ denotes the indicator function, i.e., $\InF{\mathcal{A}}{x}=1$ if $x \in \mathcal{A}$ and $\InF{\mathcal{A}}{x}=0$ otherwise.
Since in practice $E$ is finite, we further denote by $q^E_k \, p^E_k$ the system dynamics obtained by evaluating the right-hand side of \eqref{eq:it:ens_avg}, but with $E$ fixed, i.e., without taking the limit.
We notice from the definition of $\statesY$ that $|\statesY|$ is so large, even for moderate $N$, that $q^E_k \, p^E_k$ would converge to $q_k \, p_k$ only for very large values of $E$, rendering the required computations infeasible.
To overcome this problem, we consider the reduced state space $\statesYnorm = \mathcal{L}$ and map any $\vec{Y}[k] \in \statesY$ onto $\statesYnorm$ as follows
\scaleAlign
\begin{equation}%TODO: SL: Consider introducing the following function as tilde{InF}
    \InFt{\tilde{y}}{\vec{Y}[k]} = \begin{cases}
        1, &\!\!\!\textrm{ if } \vec{Y}_l[k] = \tilde{y} \textrm{ for any } 1 \leq l \leq N_Y[k],\\
        0, &\!\!\!\textrm {otherwise},
    \end{cases}
\end{equation}
where $\tilde{y} \in \statesYnorm$.
After proper normalization, we obtain
\scaleAlign
\begin{align}\label{eq:it:ens_avg_2}
    &\Pr{X_{k+1}=x_2, \tilde{Y}_{k+1}=\tilde{y}_2,X_{k}=x_1, \tilde{Y}_{k}=\tilde{y}_1}\nonumber\\
    &= \lim\limits_{E \to \infty} \frac{1}{E} \sum\limits_{e=1}^{E} \InF{\{x_1\}}{\vec{x}^e[k]} \frac{1}{N^{e}_Y[k]}\InFt{\tilde{y}_1}{\vec{Y}^e[k]}\nonumber\\
    &\quad{}\times \InF{\{x_2\}}{\vec{x}^e[k+1]} \frac{1}{N^{e}_Y[k+1]}\InFt{\tilde{y}_2}{\vec{Y}^e[k+1]},
\end{align}
where $\tilde{y}_1, \tilde{y}_2 \in \statesYnorm$, and $N^{e}_Y[k]$ denotes the number of nutrients at time step $k$ in simulation run $e$.
According to its definition, $\statesYnorm$ is equivalent to $\statesY$ for $N_Y=1$ and for $N_Y \geq 2$ it approximates states with more than one nutrient as superpositions of single-nutrient states.
We will see in the results section that the reduced state space $\statesYnorm$ still provides a meaningful characterization of the considered chemotactic system\version{}{\footnote{Further justification that the state space approximation is sufficient to clearly show the fundamental correlation between environment and system is provided in an additional evaluation section in the extended \textit{Arxiv} version of this paper \cite[Sec. IV-B]{brand2024semantic}. This section could not be included here because of space constraints.}}.
For further reference, we finally define $\tilde{q}^{E}_k \tilde{p}^{E}_k$ as the estimate of $q^E_k p^E_k$ obtained for the reduced state space $\statesYnorm$ according to \eqref{eq:it:ens_avg_2}.

We recall that, in each time step, the computational \ac{CB} introduced in the previous section senses nutrient concentration values ranging from $0$ to $9$.
Now, we seek to study within the framework of interventions, cf.~Section~\ref{sec:syn_vs_sem}, how the gathering of information from the environment impacts the ability of the \acp{CB} to adapt their behavior to the environment; as one type of interventions, we consider \acp{CB} whose sensing capability is limited to $n$ nutrients, $\mathrm{CB}_n$, where $n \in \lbrace 0, \ldots, 9 \rbrace$, i.e., $\mathrm{CB}_n$s can only sense up to $n$ nutrients.
Then, in each time step $k$, $N_n[k]$ denotes the number of nutrients sensed by a $\mathrm{CB}_n$, where $N_n[k] = \min \lbrace N^{\mathrm{o}}[k], n \rbrace$ and $\mathrm{CB}_n$ switches to tumble mode if $N_n[k] < N_n[k-1]$ or $N_n[k] = 0$, and to run mode otherwise.
Based on this definition, for each $n$, we identify $\mathrm{CB}_n$ with one intervention $\iota=n$.
For example, $\iota=0$ corresponds to $\mathrm{CB}_0$, i.e., the computational \ac{CB} always senses $0$ nutrients independent of how many nutrients actually exist in its environment; hence, $\mathrm{CB}_0$ is always in tumble mode.
On the other hand, $\iota=9$ corresponds to the default case in which the \ac{CB} can sense up to $9$ nutrients in its local environment.
In addition and to verify that the framework presented in this paper reflects some intuitive notion of adaptive and non-adaptive behavior, we introduce two more interventions $\iota=\mathrm{d}$ (dead) and $\iota=\mathrm{f}$ (fixed); intervention $\iota=\mathrm{d}$ kills the \ac{CB}, leaving it immobile and, hence, completely insensitive to the environment.
Under intervention $\iota=\mathrm{f}$, the \ac{CB} remains at a fixed location throughout the simulation, but, as long as it is alive, consumes nutrients; in that case, the \ac{CB} still interacts with the environment through the consumption of nutrients, but is not able to adapt its behavior to it.
Intuitively, both $\iota=\mathrm{d}$ and $\iota=\mathrm{f}$ reflect scenarios in which the \ac{CB} is not able to adapt to the environment and we will confirm in Section~\ref{sec:evaluation} that the semantic information framework reflects this intuition quantitatively.

Finally, we define the viability function $V[k]$ as the survival rate of the \acp{CB} at time step $k$, i.e., as the percentage of simulations for which the \ac{CB} was still alive at time step $k$.

In the following section, we will confirm that the introduced information-theoretic metrics and state definitions for computational bacterial chemotaxis provide a useful measure for the exchange of semantic information between the \ac{CB} and the environment.

\scaleSection
\section{Numerical Results}\label{sec:evaluation}
\scaleSectionBelow
In the following, we present numerical results from stochastic ensemble simulations of the computational bacterial chemotaxis framework introduced in Section~\ref{sec:abm}.
Furthermore, the reported information-theoretic measures in this section are evaluated based on the estimated system dynamics $\tilde{q}^E_k \, \tilde{p}^E_k$ as introduced in Section~\ref{sec:sem_in_chemotaxis}.
If not noted otherwise, $E = 20000$, $N=10$, $k_{\mathrm{NS}} = 2$, and $\epsilon = 0.1$, are used as default parameter values, complementing the default values already provided in Section~\ref{sec:abm}.
For each realization $e$, the initial positions of the \ac{CB} $x[0]$ and the nutrient source $s[0]$ are randomly selected on $\mathcal{L}$, respectively.

\scaleSubsection
\subsection{Viability Evaluation}
\scaleSubsectionBelow

\begin{figure}[!tbp]
\centering
  \includegraphics[width=0.95\columnwidth]{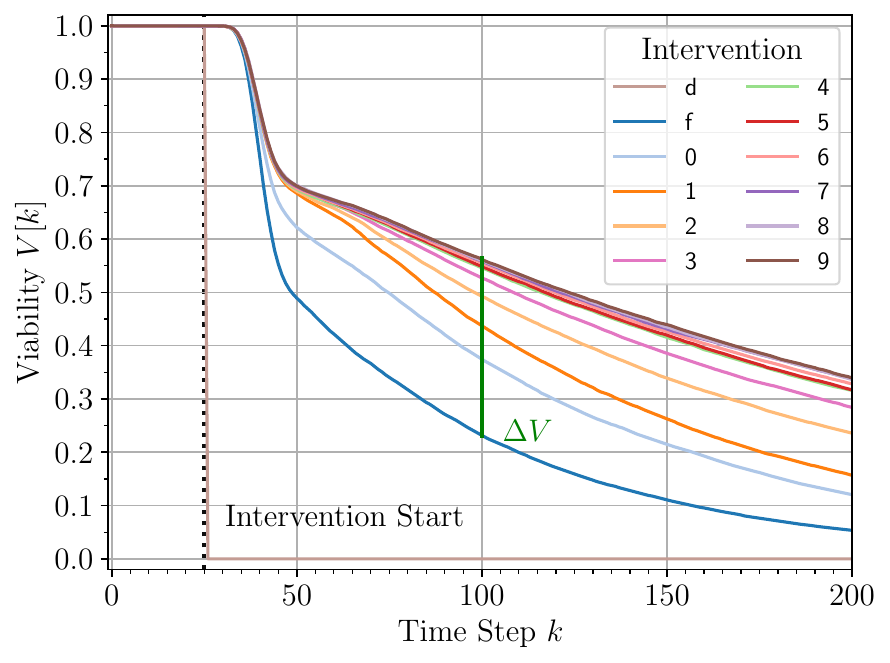}\vspace*{-0.4cm}
  \caption{Viability as a function of the simulation time steps for different interventions, which start after 25 steps. }
  \label{fig:viability}
  \vspace*{-0.3cm}
\end{figure}

Fig.~\ref{fig:viability} shows the viability of the \ac{CB} as a function of $k$ for all considered interventions $\iota \in \mathcal{I}$.
In Fig.~\ref{fig:viability}, in order to suppress the impact of initial transient dynamics (before the total number of nutrients in the environment has reached its equilibrium), the interventions are applied from $k=25$ on, i.e., the joint system-environment dynamics correspond to $q_k \, p_k$ for $k < 25$ and to $q_k \, p^{\iota}_k$ for $k \geq 25$.
We observe that the \ac{CB} with the best sensing capability, corresponding to $\iota = 9$, exhibits the largest viability while the static \ac{CB} ($\iota=\mathrm{f}$) possesses the smallest viability among all considered cases (except for $\iota=\mathrm{d}$, the dead bacterium, whose viability drops to $0$ in the moment the intervention is applied).
Fig.~\ref{fig:viability} reveals that for any $k \geq 100$ the viability is only marginally decreased for interventions $\iota=8,\ldots,4$ relative to $\iota=9$, while it decreases significantly for $\iota=3,\ldots,0$.
This observation indicates that not all the information possibly gathered by the \ac{CB} is actually relevant for its survival, motivating further investigation of the semantic information in the following.

As a reference, the difference in viability $\Delta V = V - V_{\mathrm{f}}$ between the default case, i.e., $\iota=9$, and the ``strongest'' non-lethal intervention $\iota=\mathrm{f}$, i.e., the intervention that reduces the adaptability of the \ac{CB} most while not killing it immediately, is highlighted by a green vertical line in Fig.~\ref{fig:viability}.

\version{
\scaleSubsection
\subsection{Mutual Information}
\scaleSubsectionBelow

In this section, we seek to confirm our definition of the \ac{CB}'s and the nutrients' states, respectively.
To this end, Fig.~\ref{fig:mutual_information} shows the mutual information $I_{\iota}(X_k; Y_k)$ \cite[Eq. (8.8)]{Mackay2003information} for two different parameter sets for the movement of the nutrient source, resulting in \textit{gradual} and \textit{jump} movements, and different interventions $\iota$.
We observe from Fig.~\ref{fig:mutual_information} that $I_{\iota}(X_k; Y_k)$, i.e., the dependence between the \ac{CB}'s position and the nutrients' positions, is generally largest for $\iota=9$ for both gradually and abruptly moving nutrient sources, confirming that the \ac{CB}'s sensing ability correlates with its ability to follow the moving food source.
On the other hand, we observe from Fig.~\ref{fig:mutual_information} that for $\iota=5,1$, the \ac{CB}'s ability to follow the nutrient source is decreased.
The wave-like patterns visible for $I_{\iota}(X_k; Y_k)$ for $\iota=9,5,1$ for the abruptly moving nutrient source correspond to periods of decreasing proximity between \ac{CB} and nutrient source each time $\vec{s}[k]$ changes abruptly, followed by periods of increasing proximity, when the \ac{CB} manages to realign its position with $\vec{s}[k]$.
Moreover, we observe from Fig.~\ref{fig:mutual_information} that $I_{\iota}(X_k; Y_k)$ is approximately constant for $\iota=0,\mathrm{d},\mathrm{f}$, confirming our intuition that in these cases no significant correlation between $\sysX$ and $\envY$ exists\footnote{We see that for $\iota =\mathrm{d} $, as $E$ increases, the correlation converges to $0$, which is expected, i.e., $I_{\iota =\mathrm{d}}(X_k; Y_k) \xrightarrow{E \rightarrow \infty} 0$.}.

These results confirm that the impact of the interventions on the system performance is not only reflected in the effectiveness level, i.e., in terms of the viability, as demonstrated in the previous section, but also in the correlation of the system states.
This observation confirms that the proposed state space definitions are meaningful abstractions of the actual system dynamics, motivating further investigations at the semantic level in the following section.

\begin{figure}[!tbp]
\centering
  \includegraphics[width=0.95\columnwidth]{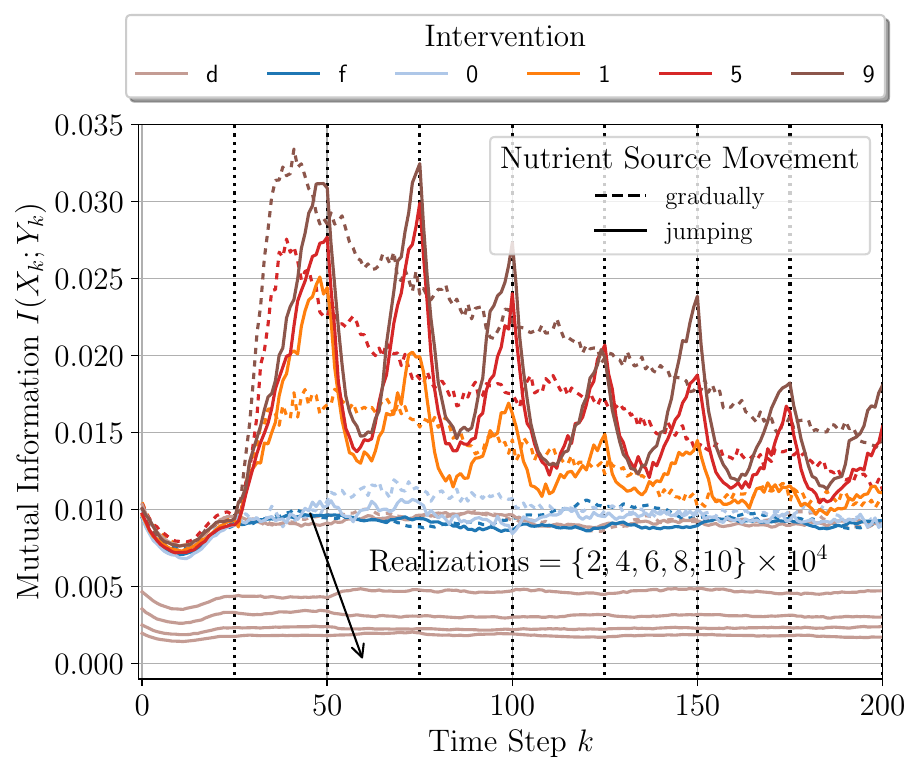}\vspace*{-0.4cm}
  \caption{The mutual information as a function of the simulation time steps for different interventions, which start after 25 steps. Nutrient source movement \textit{gradually} (dotted) (\textit{jumping} (solid)) indicates a nutrient source moving a maximum distance of $d_{\mathrm{max}} = 3$ ($d_{\mathrm{max}} = 8$) every $m = 5$ ($m = 25$) steps. The results for intervention $\iota=\mathrm{d} $ (brown) are shown for different numbers of simulation realizations $E$.}
  \label{fig:mutual_information}
  \vspace*{-0.3cm}
\end{figure}
}{}

\scaleSubsection
\subsection{Viability vs. Transfer Entropy}\label{subsec:viability_transfer}
\scaleSubsectionBelow

\begin{figure}[!tbp]
\centering
  \includegraphics[width=0.95\columnwidth]{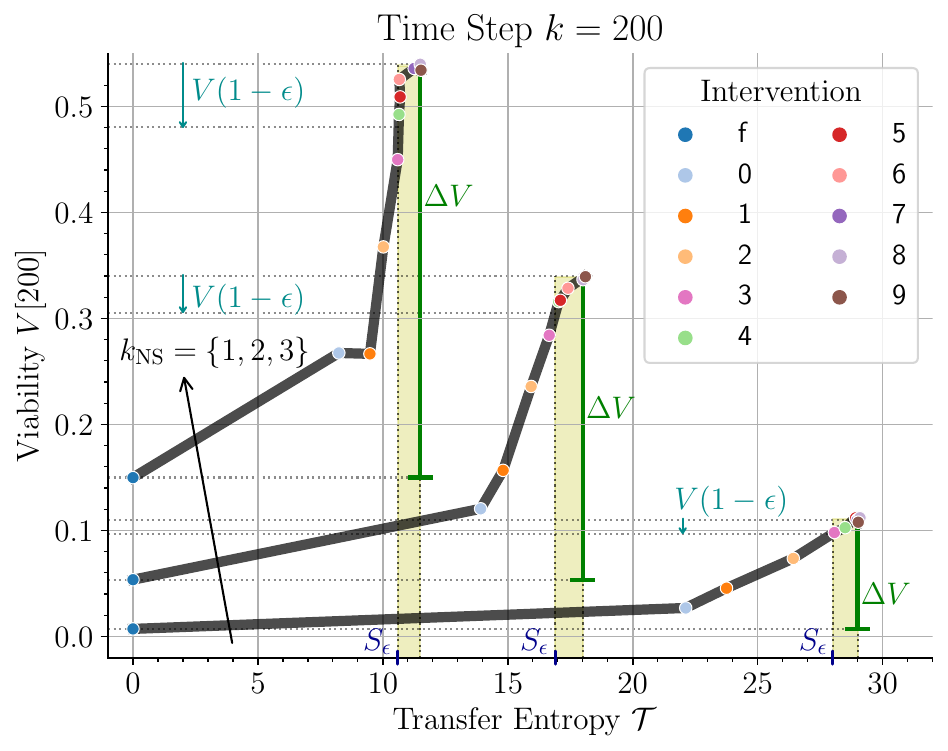}\vspace*{-0.4cm}
  \caption{Transfer entropy vs. viability at time step $k=200$ for three different nutrient production rates and different interventions. The shaded areas highlight transfer entropies of interventions resulting in (nearly) constant viability values characterized by $\epsilon$, cf. \Equation{eq:it:sem_inf}; the cyan arrows indicate the corresponding viability limits.}
  \label{fig:transfer_entropy}
  \vspace*{-0.3cm}
\end{figure}

Fig.~\ref{fig:transfer_entropy} shows the viability and the transfer entropy $\mathcal{T}_{\iota}[k]$ at $k=200$ for different nutrient production rates $k_{\mathrm{NS}}$ under different interventions.
First, we observe from Fig.~\ref{fig:transfer_entropy} that the viability is in general the higher, the more nutrients are produced. This is intuitive as more nutrients increase the chance that the \ac{CB}'s position coincides with the position of a nutrient, increasing the \ac{CB}'s chances of survival, cf. \Section{sec:bacteria}.
Next, we observe that for all considered $k_{\mathrm{NS}}$, the transfer entropy is positively correlated with the viability (except for some slight fluctuations due to the stochasticity of the numerical simulations, which are visible for $\iota = 9$); this observation confirms that the transfer entropy provides a information measure that is directly related to the ability of the \ac{CB} to fulfill its goal (effectiveness level) in a meaningful way.

Fig.~\ref{fig:transfer_entropy} also shows the observed semantic information $S_{\epsilon}$ as the minimum transfer entropy required to achieve $(1-\epsilon)\cdot 100$ percent of the maximum achievable viability $V_{9}$, where the difference between $\mathcal{T}_{9}$ and $S_{\epsilon}$ is depicted by the yellow shaded area.
We observe that for $k_{\mathrm{NS}}=1$ the \ac{CB} requires only $3$ receptors to extract all relevant information from the environment, while it requires $4$ receptors for $k_{\mathrm{NS}}=2,3$.
This observation is again intuitive; in case more nutrients are present overall, the nutrients' average local concentration at the \ac{CB} is also higher, such that the \ac{CB}'s receptors would saturate easily if there are only few of them, decreasing the \ac{CB}'s ability to detect concentration differences and follow the nutrient gradient.

Finally, we observe from Fig.~\ref{fig:transfer_entropy} that in the most challenging environment ($k_{\mathrm{NS}}=1$), the viability of the mobile ($\iota = 0,1,\ldots,9$) \ac{CB} scales approximately {\em linearly} with the transfer entropy, while this relation becomes more and more S-shaped as $k_{\mathrm{NS}}$ increases.
From this analysis, we draw the important conclusion that increased adaptability (higher transfer entropy) is directly reflected in increased fitness (viability) under challenging environmental conditions, whereas under less challenging conditions the fitness does not increase significantly further given that the adaptability exceeds some threshold value. This proves the usefulness of the proposed semantic information measure for the design of future \ac{MC} systems, especially when the system operates in and interacts with a dynamic environment.

\scaleSection
\section{Conclusion}\label{sec:conclusion}
\scaleSectionBelow
In this paper, we have utilized the semantic information framework originally proposed in \cite{kolchinsky2018semantic} to obtain an information-theoretic characterization of the considered computational bacterial chemotaxis model.
Our results demonstrate for the first time that the adaptability of the \ac{CB} is accurately reflected by the proposed framework and correlates with its ability to survive in a dynamically changing environment.
One of the considered metrics, the observed semantic information, reveals by how much the information acquisition of the \ac{CB} can be intervened (interfered with) without seriously reducing its chances of survival.
In this regard, it provides a general, yet very practical, framework for quantifying the impact of information transmission on the ability of the considered system to achieve its goal.
Since many applications envisioned for \ac{MC} share common features with the considered chemotaxis model, we believe that the presented semantic information-theoretic framework can be utilized for the analysis and goal-oriented design of future \ac{MC} systems.

\bibliographystyle{IEEEtran}
\bibliography{literature}

\begin{thebibliography}{10}
\providecommand{\url}[1]{#1}
\csname url@samestyle\endcsname
\providecommand{\newblock}{\relax}
\providecommand{\bibinfo}[2]{#2}
\providecommand{\BIBentrySTDinterwordspacing}{\spaceskip=0pt\relax}
\providecommand{\BIBentryALTinterwordstretchfactor}{4}
\providecommand{\BIBentryALTinterwordspacing}{\spaceskip=\fontdimen2\font plus
\BIBentryALTinterwordstretchfactor\fontdimen3\font minus
  \fontdimen4\font\relax}
\providecommand{\BIBforeignlanguage}[2]{{%
\expandafter\ifx\csname l@#1\endcsname\relax
\typeout{** WARNING: IEEEtran.bst: No hyphenation pattern has been}%
\typeout{** loaded for the language `#1'. Using the pattern for}%
\typeout{** the default language instead.}%
\else
\language=\csname l@#1\endcsname
\fi
#2}}
\providecommand{\BIBdecl}{\relax}
\BIBdecl

\bibitem{weaver1953recent}
W.~Weaver, ``Recent contributions to the mathematical theory of
  communication,'' \emph{ETC: A Review of General Semantics}, vol.~10, no.~4,
  pp. 261--281, 1953.

\bibitem{shi2021semantic}
G.~Shi, Y.~Xiao, Y.~Li, and X.~Xie, ``From semantic communication to
  semantic-aware networking: {Model}, architecture, and open problems,''
  \emph{IEEE Commun. Mag.}, vol.~59, no.~8, pp. 44--50, Aug. 2021.

\bibitem{yang2022semantic}
W.~Yang \emph{et~al.}, ``Semantic communications for future internet:
  {Fundamentals}, applications, and challenges,'' \emph{IEEE Commun. Surv.
  Tutor.}, vol.~25, no.~1, pp. 213--250, Nov. 2022.

\bibitem{strinati20216g}
E.~C. Strinati and S.~Barbarossa, ``{6G} networks: {Beyond} {Shannon} towards
  semantic and goal-oriented communications,'' \emph{Comput. Netw.}, vol. 190,
  p. 107930, Mar. 2021.

\bibitem{akyildiz2015internet}
I.~F. Akyildiz, M.~Pierobon, S.~Balasubramaniam, and Y.~Koucheryavy, ``The
  {Internet} of {Bio-Nano Things},'' \emph{IEEE Commun. Mag.}, vol.~53, no.~3,
  pp. 32--40, Mar. 2015.

\bibitem{ruzzante2023synthetic}
B.~Ruzzante, L.~Del~Moro, M.~Magarini, and P.~Stano, ``Synthetic cells extract
  semantic information from their environment,'' \emph{Trans. Mol. Biol. Multi
  Scale Commun.}, vol.~9, no.~1, pp. 23--27, Feb. 2023.

\bibitem{barker2023metric}
T.~S. Barker, P.~J. Thomas, and M.~Pierobon, ``A metric to quantify subjective
  information in biological gradient sensing,'' in \emph{Proc. IEEE Global
  Commun. Conf.}, Dec. 2023.

\bibitem{sowinski2023semantic}
D.~R. Sowinski \emph{et~al.}, ``Semantic information in a model of resource
  gathering agents,'' \emph{PRX Life}, vol.~1, p. 023003, Oct. 2023.

\bibitem{kolchinsky2018semantic}
A.~Kolchinsky and D.~H. Wolpert, ``Semantic information, autonomous agency and
  non-equilibrium statistical physics,'' \emph{Interface Focus}, vol.~8, no.~6,
  p. 20180041, Oct. 2018.

\bibitem{berg2004coli}
H.~C. Berg, \emph{E. Coli in Motion}.\hskip 1em plus 0.5em minus 0.4em\relax
  New York, USA: Springer, 2004.

\bibitem{nagarajan2022agent}
K.~Nagarajan, C.~Ni, and T.~Lu, ``Agent-based modeling of microbial
  communities,'' \emph{ACS Synth. Biol.}, vol.~11, no.~11, pp. 3564--3574, Oct.
  2022.

\bibitem{abar2017agent}
S.~Abar, G.~K. Theodoropoulos, P.~Lemarinier, and G.~M. O’Hare, ``Agent based
  modelling and simulation tools: {A} review of the state-of-art software,''
  \emph{Computer Science Review}, vol.~24, pp. 13--33, May 2017.

\bibitem{hibbing2010bacterial}
M.~E. Hibbing, C.~Fuqua, M.~R. Parsek, and S.~B. Peterson, ``Bacterial
  competition: {Surviving} and thriving in the microbial jungle,'' \emph{Nat.
  Rev. Microbiol.}, vol.~8, no.~1, pp. 15--25, May 2010.

\bibitem{gosztolai2020cellular}
A.~Gosztolai and M.~Barahona, ``Cellular memory enhances bacterial chemotactic
  navigation in rugged environments,'' \emph{Commun. Phys.}, vol.~3, no.~1,
  p.~47, Mar. 2020.

\bibitem{tjalma2023trade}
A.~J. Tjalma, V.~Galstyan, J.~Goedhart, L.~Slim, N.~B. Becker, and P.~R.
  Ten~Wolde, ``Trade-offs between cost and information in cellular
  prediction,'' \emph{Proc. Natl. Acad. Sci.}, vol. 120, no.~41, p.
  e2303078120, Oct. 2023.

\bibitem{rode2023chemotactic}
J.~Rode, M.~Novak, and B.~M. Friedrich, ``Chemotactic agents combining spatial
  and temporal gradient-sensing boost spatial comparison if they are large,
  slow, and less persistent,'' \emph{bioRxiv version: 2023.10.14.562229}, 2023.

\bibitem{vergassola2007infotaxis}
M.~Vergassola, E.~Villermaux, and B.~I. Shraiman, ``{‘Infotaxis’} as a
  strategy for searching without gradients,'' \emph{Nature}, vol. 445, no.
  7126, pp. 406--409, Jan. 2007.

\bibitem{Mackay2003information}
D.~J. MacKay, \emph{Information Theory, Inference and Learning Algorithms},
  1st~ed.\hskip 1em plus 0.5em minus 0.4em\relax Cambridge University Press,
  2003.

\end{thebibliography}
\scaleSectionBelow

\end{document}